\documentclass[prx,twocolumn,floatfix,superscriptaddress,longbibliography]{revtex4-1}

\usepackage{amssymb,amsmath,amstext}
\usepackage{graphicx}
\usepackage{epstopdf}
\usepackage{color}
\usepackage{bm}
\usepackage{appendix}
\usepackage[T1]{fontenc}
\usepackage{bbold}
\usepackage{bbm}
\usepackage{latexsym}
\usepackage{MnSymbol}
\usepackage[colorlinks=true,citecolor=blue,linkcolor=magenta]{hyperref}

\begin{document}

\title{
Classical surrogate simulation of quantum systems with LOWESA
}

\author{Manuel S. Rudolph}
\email{manuel.rudolph@epfl.ch}
\affiliation{Institute of Physics, Ecole Polytechnique F\'{e}d\'{e}rale de Lausanne (EPFL), CH-1015 Lausanne, Switzerland}
\affiliation{Theoretical Division, Los Alamos National Laboratory, Los Alamos, New Mexico 87545, USA}

\author{Enrico Fontana}
\affiliation{Department of Computer and Information Sciences, University of Strathclyde, 26 Richmond Street, Glasgow G1 1XH, UK}
\affiliation{Quantinuum, Terrington House, 13-15 Hills Road, Cambridge CB2 1NL, UK}
\affiliation{National Physical Laboratory, Hampton Road, Teddington TW11 0LW, UK}
\affiliation{Global Technologies Applied Research, JPMorgan Chase, 25 Bank St, London E14 5JP, UK}

\author{Zo\"{e} Holmes}
\affiliation{Institute of Physics, Ecole Polytechnique F\'{e}d\'{e}rale de Lausanne (EPFL), CH-1015 Lausanne, Switzerland}

\author{Lukasz Cincio}
\affiliation{Theoretical Division, Los Alamos National Laboratory, Los Alamos, New Mexico 87545, USA}

\begin{abstract}
We introduce LOWESA as a classical algorithm for faithfully simulating quantum systems via a classically constructed surrogate expectation landscape. After an initial overhead to build the surrogate landscape, one can rapidly study entire families of Hamiltonians, initial states and target observables. As a case study, we simulate the 127-qubit transverse-field Ising quantum system on a heavy-hexagon lattice with up to 20 Trotter steps which was recently presented in Nature \textbf{618}, 500-505 (2023). Specifically, we approximately reconstruct (in minutes to hours on a laptop) the entire expectation landscape spanned by the heavy-hex Ising model. The expectation of a given observable can then be evaluated at different parameter values, i.e. with different onsite magnetic fields and coupling strengths, in fractions of a second on a laptop. This highlights that LOWESA can attain state-of-the-art performance in quantum simulation tasks, with the potential to become the algorithm of choice for scanning a wide range of systems quickly.
\end{abstract}
\maketitle

\section{Introduction}
It is widely believed that quantum hardware, such as analog quantum simulators or digital quantum computers, will be able to tackle certain tasks \textit{better} than classical hardware. While usually the focus is on total runtime, quantum algorithms could exceed classical capabilities on a wider variety of measures, such as financial cost, energy efficiency or simply reliability. On the flip side, there exist a broad spectrum of classical methods that have their own particular niches where they excel over other state-of-the-art classical or quantum methods. Examples include tensor network (TN) methods~\cite{orus2014practical, orus2019tensor, ayral2023density, pan2022simulation}, neural quantum states~\cite{carleo2017solving,torlai2018neural,schmitt2020quantum}, and path-based Schrödinger~\cite{bernstein1997quantum,markov2018closerfarther,haner2017petabyte} or Heisenberg-picture~\cite{gottesman1998heisenberg, rall2019simulation} propagation. These classical simulation methods commonly focus on approximating the ideal solutions or dynamics that are expected given the task at hand. 

Recently, the low-weight simulation algorithm \textsc{LOWESA}~\cite{fontana2023lowesa} was introduced to approximate expectation values of Pauli-observables when the quantum circuit is affected by noise. Under the assumption of single-qubit Pauli noise channels, this algorithm is provably efficient (i.e. polynomial time) in both the number of qubits and circuit depth when the error rate is kept constant. This highlights that under these conditions, quantum algorithms executed on noisy quantum hardware cannot be expected to exhibit an exponential advantage. However, there are no guarantees for this algorithm to accurately reproduce the ideal expectation values in the absence of noise. This is of course expected due to the complexity of simulating quantum systems, and is a property that is shared by \textit{all} classical simulation methods.

In the noiseless case the main strength of \textsc{LOWESA}, over most other simulation methods, is that the algorithm reconstructs a classical representation of the full expectation landscape that is spanned by the quantum circuit parameters. More precisely, LOWESA classically constructs a \textit{surrogate} landscape~\cite{schreiber2022classical,jerbi2023shadows,landman2022classically} without requiring any reference expectation estimations, which can then be evaluated not just efficiently, but truly quickly. In the case of Variational Quantum Algorithms (VQAs)~\cite{cerezo2020variationalreview}, this implies that one could optimize the parametrized quantum circuit fully classically on the approximated landscape. 

The possibilities for simulating physical quantum systems are equally intriguing. For example, if simulating Hamiltonian dynamics using Trotter-circuits~\cite{trotter1959on,lloyd1996universal, sornborger1999higher}, the parameters of the circuits now include the coefficients of the individual Hamiltonian terms. Thus, with a single surrogate landscape, one can simulate a family of Hamiltonians that share the same operators. An additional strength of LOWESA is that its performance is not directly limited by the entangling topology of the employed quantum circuit.

\begin{figure*}[t]
    \centering
    \includegraphics[width=0.95\linewidth]{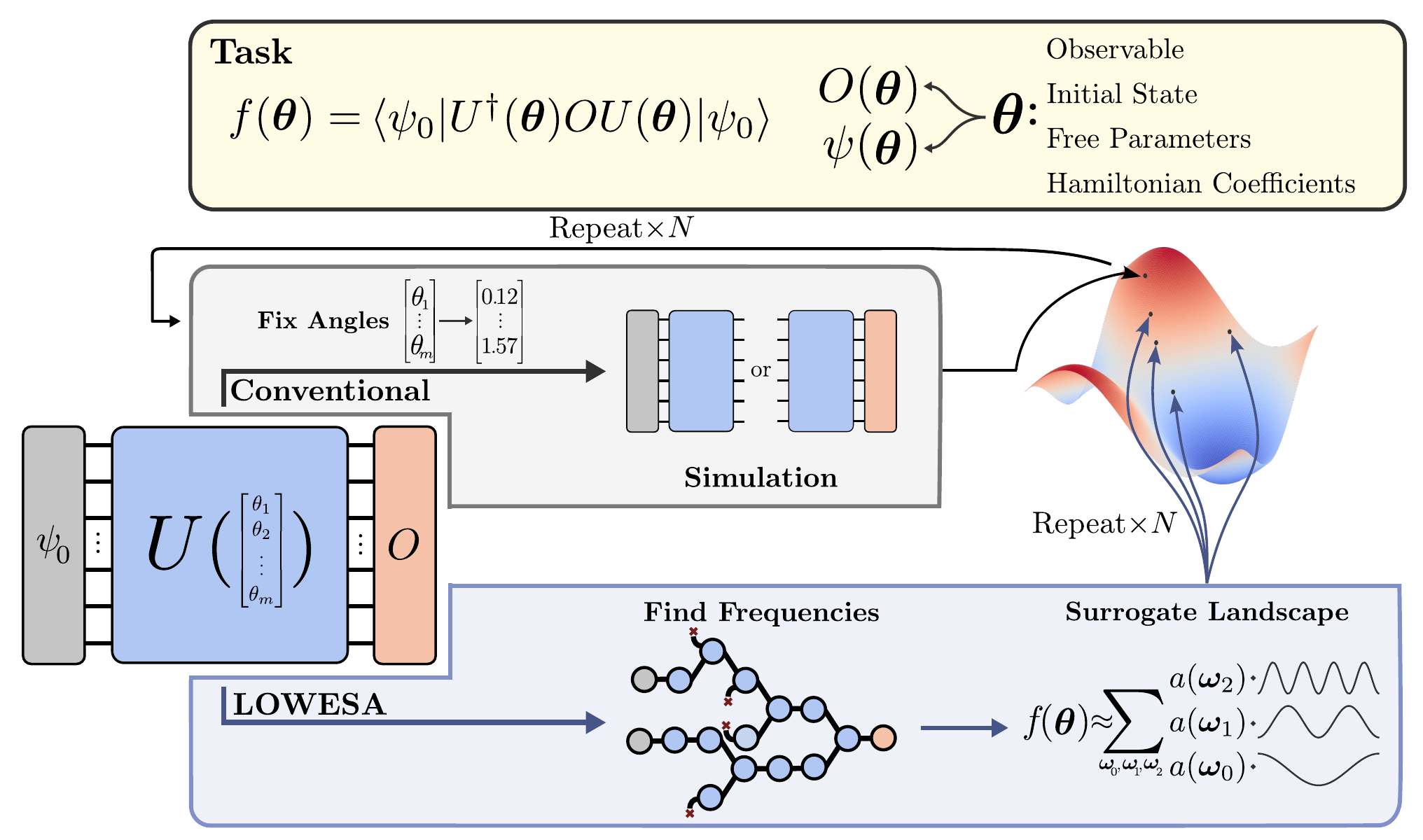}
    \caption{\textbf{Framework schematic of LOWESA for classically constructing surrogate expectation landscapes.} 
    The task is to simulate the expectation value of an observable $O$ for an initial state $\psi_0$ evolved under a unitary quantum circuit $U$ with parameters $\bm\theta$. Typically, such parameters are only introduced in the context of Variational Quantum Algorithms. However, depending on the particular task, they can also be viewed as parametrizing a change of the observable or the initial state. Another possibility is that they can indirectly be defined by the coefficients of a Hamiltonian for the purposes of studying time evolution dynamics. While simulation methods conventionally assign numerical values to the parameters $\bm\theta$ before simulation, LOWESA instead aims to build a classical surrogate of the entire expectation landscape $f(\bm\theta)$. The initial overhead of this approach may be high, but re-evaluation of the surrogate landscape for different parameter values can be exceptionally fast.}
    \label{fig:schematic}
\end{figure*}

With the aim of demonstrating the practical utility of quantum computers, Ref.~\cite{kim2023evidence} presented state-of-the-art experiments for gate-based devices using 127 superconducting qubits. The task was to estimate certain Pauli expectation values of states evolved under a transverse-field Ising (TFI) Hamiltonian for 5, 6 or 20 Trotter steps and with varying on-site magnetization. The connectivity of the TFI Hamiltonian was chosen to reflect the 2D topology of the quantum device. Using sophisticated noise-characterization techniques~\cite{berg2022probabilistic,bennett1996purification,knill2004fault} and zero-noise extrapolation (ZNE)~\cite{temme2017error,kandala2018error,giurgica2020digital}, the authors were able to obtain expectations that generally agreed with the true values in the cases where exact calculation was possible. In the cases where such verification was not possible, the authors argued that the experimentally observed trends were more plausible than those obtained with the employed tensor network methods, which had struggled throughout their work to reproduce the exact solutions. However, in a quick turn of events, three works \cite{tindall2023efficient,kechedzhi2023effective,beguvsic2023fast}, and later three more~\cite{torre2023dissipative, liao2023simulation, beguvsic2023converged} appeared that were able to classically reproduce expectation values of similar or better quality, and in less time, using distinct classical simulation techniques.

In this work, we showcase the noise-free version of the LOWESA algorithm for the example of the TFI model studied in Ref.~\cite{kim2023evidence}. To do so, we construct a surrogate for the entire expectation landscape spanned by the circuit parameters, which can be done in minutes to hours (depending on the task) on a single laptop. The surrogate can then be evaluated at different parameter values in fractions of a second to reproduce the results of Ref.~\cite{kim2023evidence} to high accuracy and with many more evaluations than Refs.~\cite{tindall2023efficient,kechedzhi2023effective,beguvsic2023fast,torre2023dissipative,liao2023simulation, beguvsic2023converged}. See Fig.~\ref{fig:schematic} for a schematic depiction of the algorithm. Interestingly, with our approach, we find that two of the expectation curves can be reproduced to high accuracy with simple trigonometric functions, namely $\sin^{25}$ and $\sin^{34}$. 

To highlight the advantages of LOWESA, we generate additional expectation curves and high-resolution expectation surfaces with low computational cost, establishing that the algorithm has a non-neglible initial overhead, but can then freely be used to probe global characteristics using the surrogate landscape. Furthermore, we go beyond the case study in Ref.~\cite{kim2023evidence} by employing quantum circuits with more than double the number of parameters, which allows us to highlight the full potential of using LOWESA for quantum simulation tasks. 

Finally, we acknowledge that our algorithm, analogous to all other classical methods that aim to reproduce exact quantum dynamics, has no guarantee to produce high-quality results within a fixed computational budget. We merely highlight its strengths in the milieu of varied and specialized classical algorithms, and advocate for a constructive interplay between quantum and classical methods to produce reliable results at large scales that can be trusted.

\section{The LOWESA algorithm}\label{sec:LOWESA}
LOWESA is a classical simulation algorithm to reconstruct the parametrized expectation landscape $f(\bm\theta)$ formed by the expression
\begin{equation}\label{eq:cost_function}
    f(\bm\theta) = \langle 0| U^\dagger(\bm\theta) O U(\bm\theta)|0\rangle,
\end{equation}
where $O$ is a normalized Pauli operator and $U(\bm\theta)$ is a quantum circuit parametrized by the vector of rotation angles $\bm\theta$.
LOWESA assumes a circuit structure composed of arbitrary Clifford operations $C_i$  and $m$ parameterized single-qubit Pauli Z-rotation gates $Rz(\theta_i)$, i.e.,
\begin{equation}\label{eq:lowesa_circuit}
    U(\bm\theta) = \prod_{i=1}^{m}\big(C_i \cdot Rz(\theta_i)\big)\cdot C_0 \, .
\end{equation}
Any quantum circuit can be represented in this form.

Originally, LOWESA was introduced for simulating variational quantum algorithms~\cite{fontana2023lowesa}. Here we consider its application to dynamical simulation. For a given dynamical simulation task composed of a set of initial states, Hamiltonians and measurements of interest, one can write down a parameterized quantum circuit that could be used to implement each of these simulations. 
After compiling these circuits into the Clifford + RZ structure in Eq.~\eqref{eq:lowesa_circuit}, LOWESA can be used to construct a surrogate for the expectation landscape. This surrogate can then be used to quickly simulate the evolution of any state, Hamiltonian and observable captured by the parameterized circuit. 

The LOWESA algorithm works in the Pauli Transfer Matrix (PTM) formalism~\cite{chow2012universal}, which is commonly applied in the Heisenberg-picture~\cite{gottesman1998heisenberg, rall2019simulation}, and to simulate open quantum systems~\cite{wood2011tensor}.  In this formalism, operators are defined by their decomposition in the Pauli basis. This means that observables and quantum states (via their density operator formulation) are treated entirely equivalently. 

Specifically, an $n$-qubit operator can be represented as a $4^n$-dimensional vector, where the $i$'th entry is defined as the corresponding coefficient for each Pauli operator $P_i\in\{I, X, Y, Z\}^n$ with  $|\cdot\rrangle_i = \text{Tr}[ (\cdot) P_i]$. For example, for a Pauli target operator $P_j$ we have $[|P_j\rrangle]_i = \text{Tr}[P_j P_i] = \delta_{ij}$. This implies that the representation of Pauli operators is sparse, with exactly one entry being non-zero. In contrast, quantum states are usually very dense.
For example $\rho_0 = |0\rangle\langle 0| = \left((I + Z)/2\right) ^{\otimes n}$. Thus, using the shorthand notation $|0\rrangle := |\rho_0\rrangle$, we have that
$[|0\rrangle]_i = \text{Tr}[|0\rangle\langle 0| P_i]$ has $2^n$ non-zero values.

A unitary $U$ is represented in the PTM formalism as $\textbf{U}$ where
\begin{equation}
    [\textbf{U}]_{ij} = \llangle P_i| \textbf{U} | P_j\rrangle = \text{Tr}[P_i U P_j U^\dagger] \, .
\end{equation}
The expectation function in Eq.~\eqref{eq:cost_function} with a parametrized unitary and a target operator $O$ represented in Pauli basis can consequently be written as
\begin{equation}\label{eq:ptm_cost_function}
    f_\text{PTM}(\bm\theta) = \llangle O | \textbf{U}(\bm\theta) | 0\rrangle = \llangle 0 | \textbf{U}^\dagger(\bm\theta) | O\rrangle \,.
\end{equation}
Since Pauli operators are sparse in the PTM formalism, it is generally easier to start the simulation with the final measurement operator $ | O\rrangle$ and propagate it backwards through the quantum circuit $\textbf{U}^\dagger(\bm\theta)$, and then calculate the overlap with the $|0\rrangle$ state. Thus LOWESA adopts this approach. 

As described in Eq.~\eqref{eq:lowesa_circuit}, the algorithm assumes that the circuit $U(\bm\theta)$ has been decomposed into Clifford operations and parametrized single-qubit Z-rotations. Clifford operations are by definition those that map one $n$-qubit Pauli operator to one other Pauli operator, i.e., they are permutations of the entries of the PTM vector $|\cdot \rrangle$. In our work, we have a well-defined gate set and thus all Clifford operations are known \textit{a priori}. This allows us to pre-compute how each operation acting on $k$ qubits acts on all possible combinations of $k$-qubit Pauli operators, and store these relations as a look-up table of size $4^k$. This time-memory trade-off allows Clifford circuits to be simulated in a time that scales exclusively linear in the number of gates. For example, the Clifford circuits that represent the endpoints $\theta_h=0, \pi/2$ of the 127-qubit system in Fig.~\ref{fig:reproduce} can be calculated in less than a millisecond.

On the other hand, the PTM simulation of non-Clifford operations is exponential in the number of such operations. This is because non-Clifford gates ``mix'' entries in $|\cdot \rrangle$ and thus drastically increase the number of non-zero entries one needs to keep track of. For a relevant example, consider the parametrized single-qubit Z-rotation gate, which, in the PTM formalism, is represented by the $4\times 4$- dimensional matrix,
\begin{equation}
    \textbf{Rz}(\theta) = \begin{bmatrix}
1 & 0 & 0 & 0 \\
0 & \cos(\theta) & -\sin(\theta) & 0 \\
0 & \sin(\theta) & \cos(\theta) & 0 \\
0 & 0 & 0 & 1 
\end{bmatrix}.
\end{equation}

A common strategy in this case is to use path-based methods that decompose inefficient components (here the RZ-gate) into a sum of efficient components. This transforms one exponentially costly calculation into exponentially many efficient calculations. While still exponential in nature, path-based methods allow for finite-fidelity approximations using a non-exponential amount of computational resources. This is due to the ability to neglect some (or most) terms in the decomposition, which can lead to efficient trade-offs.
In our case, the RZ-gate admits the decomposition into three operations \begin{equation}
    \textbf{Rz}(\theta) = \textbf{D}_0 + \cos(\theta) \textbf{D}_1 + \sin(\theta) \textbf{D}_{-1} \,,
\end{equation}
with
\begin{align}
    \textbf{D}_0 &= \begin{bmatrix}
1 & 0 & 0 & 0 \\
0 & 0 & 0 & 0 \\
0 & 0 & 0 & 0 \\
0 & 0 & 0 & 1 
\end{bmatrix}
\end{align}
\begin{align}
\textbf{D}_1 &= \begin{bmatrix}
0 & 0 & 0 & 0 \\
0 & 1 & 0 & 0 \\
0 & 0 & 1 & 0 \\
0 & 0 & 0 & 0 
\end{bmatrix}
\end{align}
\begin{align}
\textbf{D}_{-1} &= \begin{bmatrix}
0 & 0 & 0 & 0 \\
0 & 0 & -1 & 0 \\
0 & 1 & 0 & 0 \\
0 & 0 & 0 & 0 
\end{bmatrix}.
\end{align}

Since in the PTM formalism the four-dimensional vector representing any single-qubit operator contains the coefficients of $I, X, Y, Z$  Pauli operators (by convention in this order), the $D_0$ operator projects any $X, Y$ contribution to zero (we call this annihilation), whereas the $D_1, D_{-1}$ operators annihilate any contribution of $I, Z$. That is, a path either does not split, which means that the $D_0$ operator is applied, or it does split into two paths, where the $D_1$ or $D_{-1}$ operator are applied to each path, respectively. On the subspace that these operators do not annihilate, they act like Clifford gates, i.e. they permute incoming vectors. Using this decomposition, a worst-case simulation of the expectation function in Eq.~\eqref{eq:ptm_cost_function} would require the calculation of $2^m$ so-called paths where every application of an RZ-gate splits the incoming operator into two, where $D_1$ is applied to one and $D_{-1}$ to the other.

By linearity, the final expectation value is a straight sum of all paths.
Specifically, for any path that is defined by its coordinate vector $\bm\omega = \{0, 1, -1\}^m$, the expectation function $f$ can be written as 
\begin{equation}\label{eq:lowesa_cost_function}
    f(\bm\theta) = \sum_{\bm\omega}\Phi_{\bm\omega}(\bm\theta) \llangle 0| \textbf{U}^\dagger_{\bm\omega}| O\rrangle \, ,
\end{equation}
where 
\begin{equation}\label{eq:trig_monomial}
    \Phi_{\bm\omega}(\bm\theta) = \prod_{i=1}^m \begin{cases} 
    1 & \mbox{if } \omega_i = 0 \\ 
    \cos(\theta_i) & \mbox{if } \omega_i = 1 \\
    \sin(\theta_i) & \mbox{if } \omega_i = -1
    \end{cases} 
\end{equation}
is a parameter-dependent trigonometric monomial, and $\llangle 0| \textbf{U}^\dagger_{\bm\omega}| O\rrangle \in \{-1, 0, 1\}$ is the expectation value of the purely-Clifford circuit $\textbf{U}^\dagger_{\bm\omega}$ for the path indexed by $\bm\omega$. Here, each RZ-gate at position $i$ is replaced by one of the operators $\textbf{D}_0, \textbf{D}_1, \textbf{D}_{-1}$ depending on the value of $\bm\omega$ at position $i$. 

From another perspective, this formulation in Eq.~\eqref{eq:lowesa_cost_function} can be understood as a Fourier-series representation of $f$, where $\Phi_{\bm\omega}(\bm\theta)$ are trigonometric basis functions, and $\llangle 0| \textbf{U}^\dagger_{\bm\omega}| O\rrangle$ are the Fourier coefficients. In this picture, $\bm\omega$ can be seen as the \textit{frequency vector}, as via Eq.~\eqref{eq:trig_monomial}, one can directly connect the 1-norm $|\bm\omega|_1$ of this vector to the frequency with which this path contributes to the landscape.
Thus, paths with many splits generally contribute highly oscillatory features, while paths with fewer splits determine the coarse-grained outlines of the landscape.

In summary, LOWESA is an algorithm to construct a surrogate function $\tilde{f}$ for the exact function in Eq.~\eqref{eq:lowesa_cost_function}, using different forms of truncation to keep the computation tractable. Due to its Fourier-series nature, this results in forming a surrogate for the entire expectation landscape defined by $f(\bm\theta)$. 

In the context of quantum simulation, there might not appear to be a concept of a landscape, as the goal is often to run a fixed quantum circuit derived from exponentiating a Hamiltonian.
If the circuit, for example, approximates time evolution under a Hamiltonian until time $t$ via the Trotter-Suzuki decomposition, the effective ``angles'' of the circuit are directly proportional to Hamiltonian coefficients and the time step $\Delta t$. However, by adapting the view of these values as parameters, one can use LOWESA to simulate time evolution under the entire family of Hamiltonians that share the same structure, while allowing complete freedom to change the coefficients or $\Delta t$ afterward. We note that, if the goal is to simulate only a single circuit instance, LOWESA in its general form is unlikely to be the best technique.

\medskip

The critical insight in Ref.~\cite{fontana2023lowesa} is that LOWESA is provably efficient at simulating quantum circuits affected by single-qubit Pauli noise, as paths with many splits, i.e. those with large 1-norm $|\bm\omega|_1$, are suppressed. In other words, high-frequency contributions to the expectation landscape are suppressed most by noise.
Here we however aim to simulate the exact expectation function $f$. In this case, the error one makes by truncating is unbounded in general-- a property shared by all classical simulation methods that aim to simulate exact quantum circuits. Nonetheless, as demonstrated in Section~\ref{sec:Results}, in many cases our algorithm can be used for sufficiently accurate simulations, even for the 127-qubit systems considered here.

Shortly before the original publication of LOWESA~\cite{fontana2023lowesa}, the authors in Ref.~\cite{nemkov2023fourier} showcased a similar algorithm and presented a study of the Fourier series properties of VQAs. Among other insights, they found that strongly-contributing frequencies clustered and generally were lower than one might expect. These findings help explain LOWESA's proficiency at simulating quantum systems at scale. Soon after the publication of the 127-qubit experiments in Ref.~\cite{kim2023evidence}, a similar approach under the name of \textit{Clifford perturbation theory} (CPT) was used in Ref.~\cite{beguvsic2023fast} to simulate single expectation values from the time-evolved state (see Ref.~\cite{beguvsic2023simulating} for more details about CPT). While sharing similarities with our line of research, neither work developed a surrogate-based algorithm for simulating the dynamics of wide families of Hamiltonians, initial states and observables.

\section{Approximation and Tailoring Techniques}\label{sec:approx_and_tailoring}
In this section, we introduce a selection of empirical truncation or computational tree-pruning techniques that we use to apply LOWESA to large-scale problems with hundreds or thousands of parametrized gates. These methods assume that finding all paths is unfeasible or unnecessary. Thus, we introduce a sense of prioritization in the tree search for contributing paths. We summarize these methods below and discuss them in more detail in Appendix~\ref{apx:convergence}.

\medskip

\emph{Maximal frequency $\ell$:} 
The simplest truncation criterion is to drop all paths with $|\bm\omega|_1 > \ell$ for a truncation parameter $\ell$. This reduces the absolutely worst-case scaling of the tree search for valid paths from $2^m$ paths to $2^\ell$. This truncation scheme is particularly justified for simulating noisy circuits since, as discussed above and as shown in Ref.~\cite{fontana2023lowesa}, high $|\bm\omega|_1$ paths are most suppressed by noise. In the noise-free case, the error one makes with any $\ell < m$ is trivially bounded by $1$. However, via the interpretation of $\bm\omega$ as the frequency of a path, truncation by $\ell$ can be seen as enforcing a low-frequency approximation of the expectation landscape.

\medskip

\emph{Truncation probability $p$:} This parameter provides a way to preemptively truncate paths based on an estimate of the likelihood that it will split more often than $\ell$. To do so, we crudely assume that the probability of splitting is uniform across circuit layers and extrapolate from the history of the number of splits so far to estimate the probability that a path will remain with $|\bm\omega|_1 \leq \ell$. Once that probability goes below $p$, we preemptively truncate that path and do not explore it any further.
This method of truncation can save a lot of computational effort by neglecting unpromising paths early, but it may introduce a bias against paths that split often early in the circuit, and generally those with frequencies close to the cutoff $\ell$. We have found that there exist values for $p$ that are \textit{small enough} to not neglect many valuable paths that would have stayed below the threshold $\ell$ but still save computational time; however, this optimal value cannot be known in general.

\medskip

\emph{Maximal operator weight $W$:} This parameter provides a way to preemptively truncate paths based on the current \textit{weight} of the back-propagating Pauli operator. The weight of an operator is defined as the number of qubits that it acts on non-trivially, i.e., with $X$, $Y$, or $Z$ (not $I$). We truncate a path early if the back-propagated operator surpasses a weight of $W$ at any stage in the circuit.
The weight of the operator can be understood as an \textit{effective lightcone per path}, which is closely related to the practice of using effective entanglement lightcones for lowly-entangling circuit layers (see Ref.~\cite{kechedzhi2023effective} for a relevant example). 
Furthermore, the weight of the operator makes it exponentially more likely that a path contains $X$ or $Y$ operators on any qubit at the end of the circuit and therefore not contribute to the landscape as $\llangle 0| \textbf{U}^\dagger_{\bm\omega}| O\rrangle = 0$. Thus, this truncation method significantly increases the odds that a path eventually contributes to the surrogate landscape formed by LOWESA. 

\begin{figure*}
    \centering
    \includegraphics[width=0.99\linewidth]{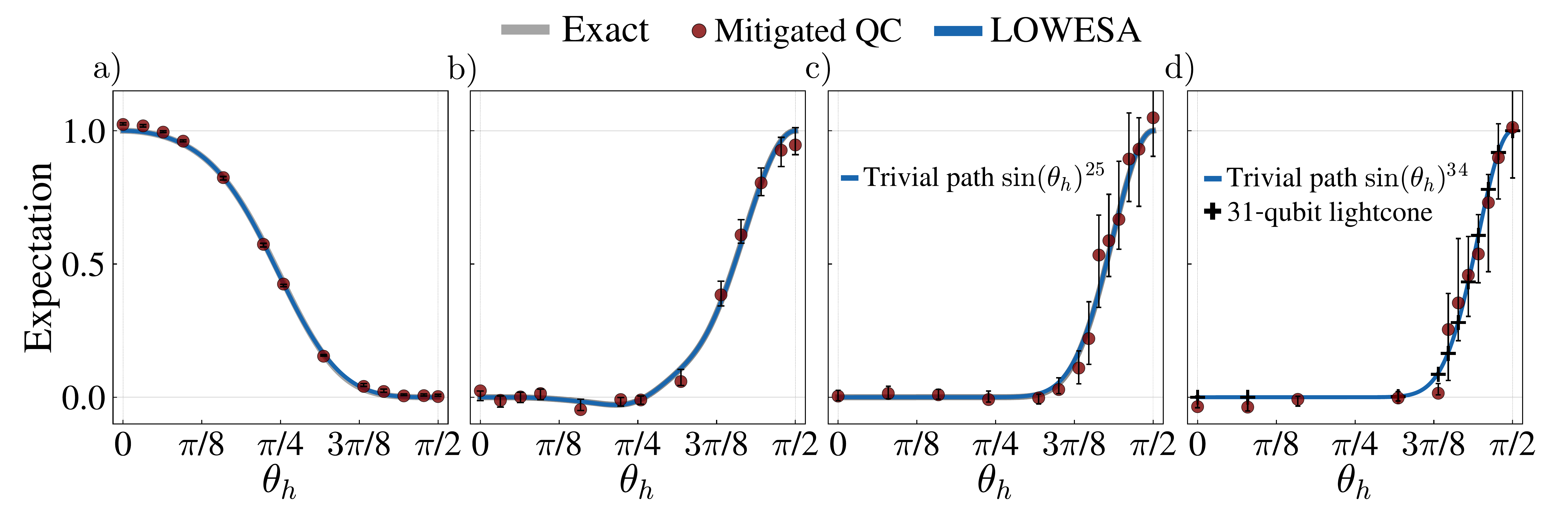}
    \caption{\textbf{Simulating 127-qubit TFI dynamics.} The observables measured correspond to Figs. 3a, 3b, 3c and 4a from Ref.~\cite{kim2023evidence}, which also provides the exact curves and the mitigated quantum computer (QC) results. The observables are the magnetization $M_z = \frac{1}{127}\sum_i\langle Z_i\rangle$, $\langle X_{13,29,31}Y_{9,30}Z_{8,12,17,28,32}\rangle$, $\langle X_{37,41,52,56,57,58,62,79}Y_{75}Z_{38,40,42,63,72,80,90,91}\rangle$, and $\langle X_{37,41,52,56,57,58,62,79}Y_{38,40,42,63,72,80,90,91}Z_{75}\rangle$, respectively. After reconstructing the expectation landscape of these observables with $L=5$ (a-c) or $L=6$ (d) trotter steps, we are able to replicate close to exact expectation curves for the correlated angle case $\theta_h^{(i)}=\theta_h$. Note that the exact curves are mostly hidden because of the high-quality approximation. In panel a) we use $\ell=40, W=8$, and in panel b) $\ell=32, p=0.05$. For this restricted 1D ``slice'' of the expectation landscape, the trivial paths corresponding to the trigonometric monomials $\sin^{25}$ and $\sin^{34}$ were sufficient to provide near-exact dynamics in panel c) and d), respectively. The 31-qubit lightcone simulations were generated with the code accompanying Ref.~\cite{kechedzhi2023effective}.}
    \label{fig:reproduce}
\end{figure*}
\medskip

\emph{Trivial paths:} Trivial paths are what we call paths that satisfy the boundary conditions of the parameter domain of interest. For the simulation tasks considered here, they tend to be the biggest contributors to the qualitative shape of the expectation landscape, i.e., among the lowest-frequency contributions, and are exceptionally easy to find. Additional paths then form corrections that eventually reveal the true dynamics, but we show that the trivial paths alone can replicate two of the expectation curves in Ref.~\cite{kim2023evidence} (and can be found in less than a millisecond).
For example, consider the case of a single RZ-gate with angle $\theta$ in an otherwise Clifford circuit. We can then verify the boundary cases of $\theta = 0$ and $\theta = \pi/2$. If any of the expectations at the boundaries is 1, we can be sure that either the path that \textit{always} goes down the $D_{1}$ or $D_{-1}$ channel (depending on whether $\theta = 0$ or $\theta = \pi/2$ has expectation 1, respectively) is valid and satisfies this boundary value because these are the only paths that can possibly be non-zero at these angles.

\medskip

\emph{Sine or cosine biasing:} Similarly to the trivial path method above, we can use information about verifiable domains of the expectation landscape to inform us which types of paths are most likely to contribute significantly. That is to say, we gain information on whether amplitudes of relevant paths are expected to have a larger contribution of $\sin$ or $\cos$ on the parameter domain of interest. For correlated angles, as is the case in Figs.~\ref{fig:reproduce} and~\ref{fig:4b_and_more}a, we have found that contributing paths tend to have a strong tendency to mostly split into $D_1$ or $D_{-1}$ in exact correspondence with the dominant trivial path.

\section{Results}\label{sec:Results}

We consider the time-evolution dynamics of a 127-spin system governed by the transverse-field Ising (TFI) Hamiltonian
\begin{equation}\label{eq:TFI}
    H = -\sum_{\langle i,j \rangle} J^{(i,j)} Z_iZ_j + \sum_i h^{(i)} X_i \,,
\end{equation}
where $X$ and $Z$ represent X- and Z-Pauli operators, respectively, and $\langle i,j \rangle$ denotes neighboring spin indices on the so-called \textit{heavy-hex} topology (also shown in Fig.~\ref{fig:magnetization}).
See Ref.~\cite{kim2023evidence} for details.

The evolution operator under this Hamiltonian for time $t$ can be written as $U(t) = e^{-i H t}$. This cannot be implemented directly on digital quantum devices due to non-commuting operators in $H$. Instead, one can use the Trotter-Suzuki decomposition~\cite{trotter1959on,lloyd1996universal, sornborger1999higher},
\begin{equation}\label{eq:trotter-circuit}
    U(\bm\theta) = \prod_{l=1}^L \left[\prod_{\langle i,j \rangle}\left( R_{ZZ}(\theta_{J}^{(i,j)})\right) \prod_i \left( R_{X}(\theta_{h}^{(i)}) \right) \right]\, ,
\end{equation}
which breaks up the evolution into $L$ discrete time steps of length $\Delta t = t/L$.
Here $R_{ZZ}(\theta)$ and $R_{X}(\theta)$ are Pauli-rotation gates with rotation angles $\theta_{J}^{(i,j)} = -J^{(i,j)} \Delta t$ and $\theta_{h}^{(i)} = h^{(i)} \Delta t$. With $\Delta t \xrightarrow[]{} 0$, this approximate time evolution circuit becomes exact. For LOWESA, the parameters $\theta$ are freely adjustable after computing the surrogate which allows us to reconstruct expectation landscapes of the TFI dynamics for arbitrary individual coupling strengths $J^{(i,j)}$, magnetic fields $h^{(i)}$ and time steps $\Delta t$.

\medskip

\emph{Reproducing the simpler experiments:}
In a first experiment, we verify that we can reconstruct the expectation values obtained on a 127-qubit quantum computer in Ref.~\cite{kim2023evidence}. For these cases, the coupling coefficients are all fixed at $J^{(i,j)} = J = -\pi/2 \; \forall \; i,j$, such that $R_{ZZ}(\theta_{J}^{(i,j)})$ are Clifford gates. The local fields are also all identical, i.e. $h^{(i)} = h \; \forall \; i$,  and are tuned in unison such that $\theta_{h}^{(i)}=\theta_{h} \in [0, \pi/2]$. Our results are shown in Fig.~\ref{fig:reproduce}. We note that panels a) to c) use a circuit with $L=5$ Trotter layers, whereas panel d) uses a circuit with an additional layer  of $R_X$ rotations, which effectively corresponds to $L=6$ (as the $R_{ZZ}$ layer acts trivially on the initial state~$|0\rangle$).

For the cases where exact verification was possible, we show that LOWESA recovers the dynamics with remarkably high precision, and is always in good agreement with the error-mitigated experimental results from Ref.~\cite{kim2023evidence}. The surrogate construction took under an hour per observable (sometimes significantly less), and then it took fractions of a second to evaluate each of the 158 values of $\theta_h$ per curve. As such, our approach is competitive with the methods in Refs.~\cite{tindall2023efficient} and~\cite{beguvsic2023fast}, which took of the order of minutes per $\theta_h$ value, for reproducing the full curves. However, our surrogate landscapes can then be used to quickly probe significantly more parameter values, including uncorrelated ones with $\theta_{h}^{(i)}$ acting independently on each qubit $i$.

\begin{figure*}
    \centering
    \includegraphics[width=0.95\linewidth]{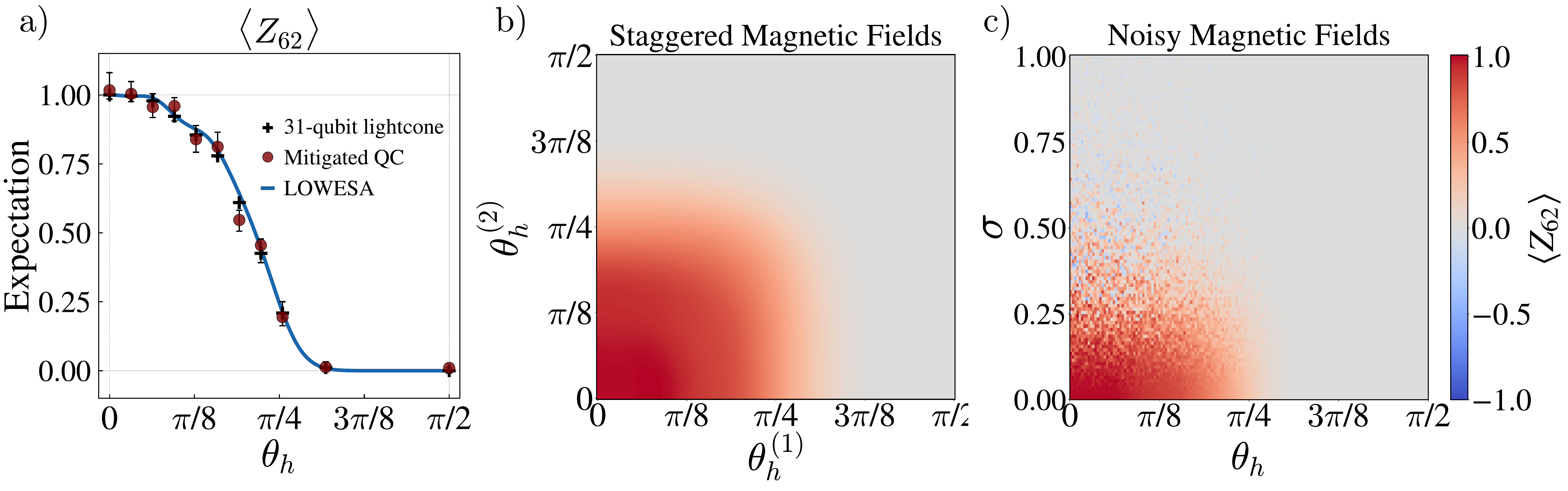}
    \caption{\textbf{Simulating 127-qubit TFI dynamics with $L=20$ Trotter steps.} Panel a) showcases the example of Fig.~4b from Ref.~\cite{kim2023evidence}, for which we are able to generate very plausible results that are broadly within error bars of the quantum hardware results and close to the approximate results demonstrated in Ref.~\cite{kechedzhi2023effective}. Since we approximately reconstructed the entire expectation landscape spanned by the Hamiltonian coefficients $h^{(i)}$ in Eq.~\eqref{eq:TFI}, we are able to generate additional $158\times 158$ pixel expectation surfaces measuring $\langle Z_{62}\rangle$ for alternating magnetic field values on the qubits (b) and for random noise on the magnetic fields $h^{(i)}$, drawn from a normal distribution with standard distribution $\sigma/\Delta t$ on each qubit (c). Both panels b) and c) contain the curve in panel a) as special case, which is along the diagonal of panel b), and the x-axis of panel c) with $\sigma=0$.}
    \label{fig:4b_and_more}
\end{figure*}

Our results in Fig.~\ref{fig:reproduce} highlight four of the approximation and tailoring techniques outlined in Sec.~\ref{sec:approx_and_tailoring}. 
Firstly, we always use a frequency truncation $\ell<m$ where $m = 635$ for panels a)-c), and $m=762$ in panel d). This is motivated by lower-frequency trigonometric functions generally being more important for revealing general trends. Furthermore, because these cases have correlated angles $\theta_h$, one can see that $\sin(\theta_h)^q\cdot\cos(\theta_h)^r$ is exponentially suppressed to zero for large integer values $q,r$ for all $\theta_h$. Thus, one would expect that only cases where $q\ll r$ or $r \ll q$ have significant amplitudes on this parameter range. Depending on the verifiable boundary cases (i.e. the magnetization values at $0$ and $\pi/2$), one may therefore additionally bias the tree search to predominantly focus on the $D_1$ or $D_{-1}$ channels. For uncorrelated parameters, these latter arguments do not hold.

Secondly, in panel a) we utilize truncation based on the weight of the back-propagated observable with $W=8$, and in panel b) we utilize a modest truncation probability of $p=0.05$. These values may not be optimal to reproduce the entire uncorrelated parameter landscape, but they suffice for the cases presented here. It is likely that there exist stronger truncations which heavily reduce computational time but still accurately approximate the exact dynamics for the chosen parameter range. However, since the optimal hyperparameter values are likely to be highly problem-specific, and for non-trivial problems one is unlikely to have access to the exact dynamics to compare to, we were not interested in fine-tuning the result with the lowest possible computational resources for these cases. 

And finally, under our approach, panels c) and d) can be reconstructed in milliseconds because the trivial paths strongly dominate the expectation values of the target operators. As mentioned in Sec.~\ref{sec:approx_and_tailoring}, from the verifiable Clifford circuit cases at $\theta_h =0$ and $\theta_h = \pi/2$, we can infer that the paths that always take the $D_{-1}$ branch, i.e., where $\Phi_{\bm\omega}$ consist only of $\sin$-functions,  must exist and have non-zero amplitudes. That is, this is the only path that can take an expectation value of $1$ at $\theta_h = \pi/2$.
Applying LOWESA to find these paths is effectively instantaneous and results in the amplitudes $\sin(\theta_h)^{25}$ and  $\sin(\theta_h)^{34}$ for panels c) and d), respectively. Additional paths would then form corrections to these trivial paths, but in this work, we did not find any additional paths with sizeable amplitudes in this parameter range, suggesting even without exact verification to compare to that the trivial reconstruction is of high quality.

\medskip

\begin{figure*}
    \centering
    \includegraphics[width=0.95\linewidth]{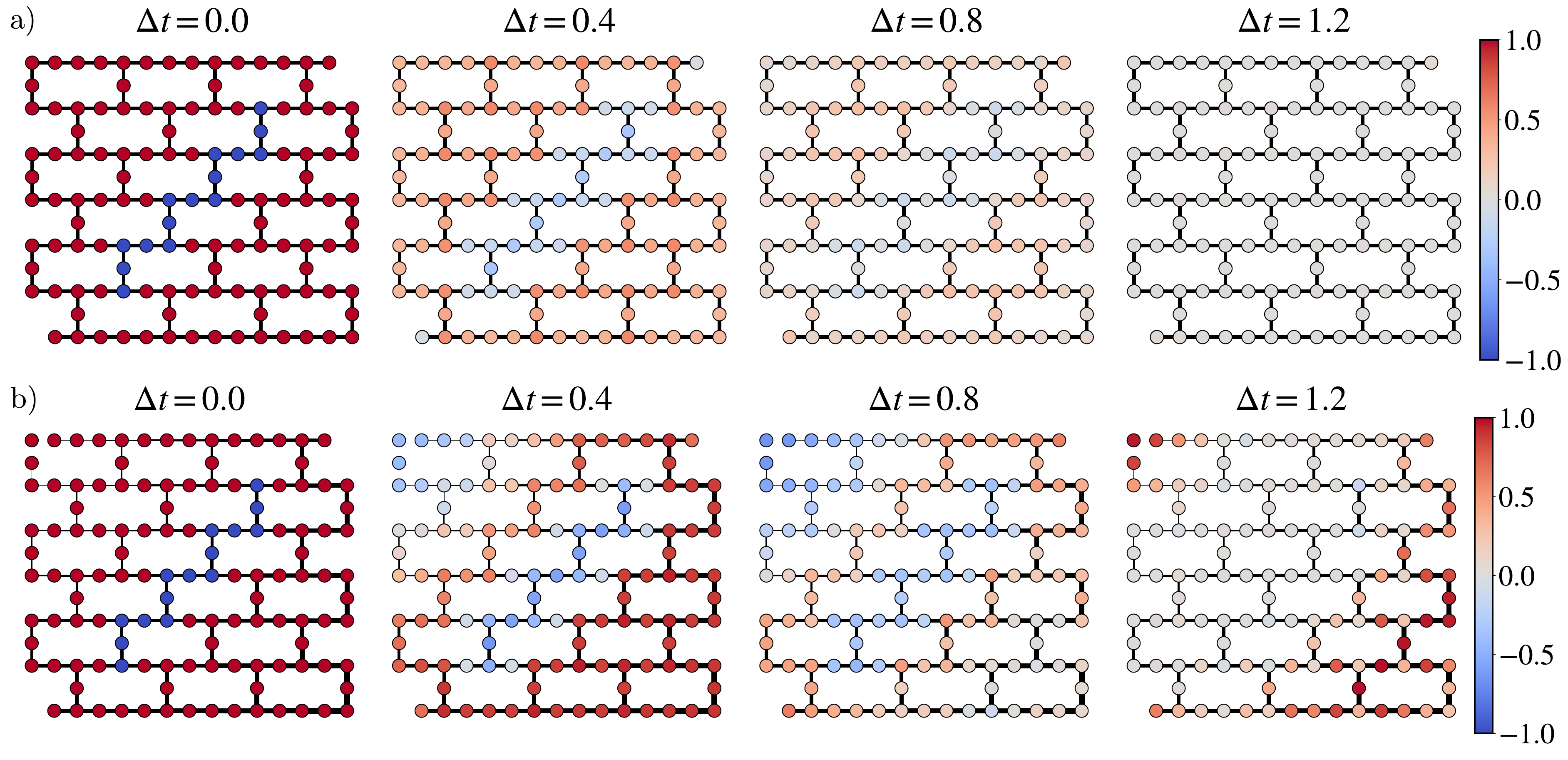}
    \caption{\textbf{Simulating 127-qubit TFI dynamics with individually controllable angles $\theta_J^{(i,j)}, \theta_h^{(i)}$.} Extending the case of Fig.~\ref{fig:reproduce}a, we now employ the circuit in Eq.~\eqref{eq:trotter-circuit} with fully flexible angles, i.e., non-Clifford entangling gates. Here we show how each single-site magnetization evolves in time given the initial condition labeled by $\Delta t = 0$, and either $J^{(i, j)}=1$ between all qubits (a), or a diagonal linear ramp (b) from $J^{(0, 1)} = 0$ at the top left qubit to $J^{(125, 126)}= -3$ at the lower right (indicated by the thickness of the connecting edges). Having reconstructed the expectation landscapes for all 127 single-qubit $Z$ operators, recalculation of a snapshot in time can be done in seconds.}
    \label{fig:magnetization}
\end{figure*}

\emph{Reproducing the hardest experiments:}
We now move on to reproducing Fig.~4b from Ref.~\cite{kim2023evidence}. Here, the circuit is significantly deeper with $L=20$ Trotter layers, such that even with a single-qubit Z-observable, the entanglement lightcone of the circuit includes all qubits. This case seems to be significantly harder to reproduce using LOWESA, mostly because we can quickly find millions of paths, many of them contributing significantly, and in turn fill up the RAM of a laptop. Using a range of moderate truncation values, we do however find trends that are generally in good agreement with the experimentally observed values, as well as with the 31-qubit effective lightcone simulations in Ref.~\cite{kechedzhi2023effective}. In Appendix~\ref{apx:convergence}, we provide a detailed study of how the expectation curves behave for different truncation parameters, and differences in convergence behavior that we expect compared to, for example, tensor network simulations. One of the curves that we found to be converging in Fig.~\ref{fig:weight-convergence} in Appendix~\ref{apx:convergence} is depicted in Fig.~\ref{fig:4b_and_more}a.

Having built the surrogate landscape and verified its quality against other methods, we can now depart from the correlated-angle case. We highlight this in Fig.~\ref{fig:4b_and_more}b and~c, which contain Fig.~\ref{fig:4b_and_more}a (i.e. Fig.~4b from Ref.~\cite{kim2023evidence}) as a special case. First, in panel b) we show a 2D expectation surface spanned by two parameters $\theta_{h}^{(1)}$ and $\theta_{h}^{(2)}$, which control all odd and all even qubit indices, respectively. The diagonal, $\theta_{h}^{(1)}=\theta_{h}^{(2)}$, contains the data shown in panel a). Second, we allow each of the $2540$ single-qubit RZ-gates to vary independently by adding Gaussian random noise with zero mean and standard deviation $\sigma /\Delta t$ to the fields $h^{(i)}$ on each qubit $i$. 

We stress that the surface plots shown in Fig.~\ref{fig:4b_and_more} are only illustrative examples: there are a myriad of different cuts of the expectation landscape one could study with our surrogate. Similarly, with this surrogate one could simulate any product initial state with $R_X$ rotations. Or, more generally, by shorterning the circuit depth one could simulate entangled initial states and more complex measurements.  
This demonstrates the flexibility of LOWESA- one constructs the surrogate, and then can flexibly explore different simulation regimes afterward. 

We note that most other classical simulation techniques would take significantly longer than the approximately two hours it took us to evaluate the $158\times 158$ surface on a 128 CPU-core machine. This wall time advantage may potentially even hold when including the construction of the high-quality surrogate landscape (see convergence in Fig.~\ref{fig:weight-convergence}), which took on the order of one day. Furthermore,  with every additional surface or initial state simulated via the surrogate, the time saved using LOWESA, as compared to other established approaches, only increases.

Recent results using tensor network techniques~\cite{anand2023classical,liao2023simulation} found that the true expectation values in the middle section of the curve in Fig.~\ref{fig:4b_and_more}a may be slightly larger than those predicted by the non-tensor network methods~\cite{beguvsic2023fast,kechedzhi2023effective}, those obtained on noisy quantum hardware~\cite{kim2023evidence}, or recent large-scale simulations in Ref.~\cite{beguvsic2023converged}, which were claimed to be converged. This raises the interesting topic of verification beyond the exactly verifiable regime. Each simulation method likely has its niche where they are most accurate, and we show that LOWESA can contribute to that joint effort of simulating quantum systems. In particular, other simulation methods can be used to verify certain corners of LOWESA's expectation landscape so that it can be trusted when quickly observing novel sections.

\medskip

\emph{Beyond Clifford entangling gates:} To go beyond prior results on this TFI model, we now employ Trotter time evolution circuits where all parameters $\theta_{J}^{(i,j)}$ and $\theta_{h}^{(i)}$ are free parameters. While this makes the simulation significantly harder by more than doubling the number of free parameters and making all gates in Eq.~\eqref{eq:trotter-circuit} non-Clifford, it gives us complete control over all Hamiltonian coefficients in Eq.~\eqref{eq:TFI} as well as $\Delta t$. Previously, this value was not independently tunable, due to the constraint that $\theta_{J} = -J \Delta t = -\pi/2$ had to remain constant to make the entangling gate Clifford. 

Applying this circuit to studying the magnetization \mbox{$M_z = \frac{1}{127}\sum_i \langle Z_i\rangle$}, as in Fig.~\ref{fig:reproduce}a, we can investigate how each qubit's magnetization evolves in time depending on the initial condition and coupling strengths $J^{(i,j)}$. While LOWESA generically assumes the all-zero initial state $|0\rangle^{\otimes n}$, the existing first layer of RX-rotations can naturally be used to create a boundary wall with individual qubits starting in the $|1\rangle$ state. Furthermore, we can impose a gradient in the coupling strengths on the topology of the qubits, and then evolve the system in time by changing $\Delta t$. These are all capabilities of LOWESA that are enabled by the fast re-evaluation of the reconstructed surrogate landscape and highlight how circuit parameters can be interpreted in several ways with distinct applications.

Fig.~\ref{fig:magnetization} depicts the single-qubit magnetization on the 127-qubit heavy-hex topology evolving in time for $h^{(i)}= h = 1$ and $J^{(i,j)}=J=1$ in panel a) and for a linear ramp in the couplings from the top left to the lower right in panel b). This results in $J^{(0, 1)} = 0$ at the top left qubit to $J^{(125, 126)}= -3$ at the lower right. While the homogeneous coupling case exhibits a steady decay of magnetization over time, the non-homogeneous case displays more intricate dynamics where large couplings initially block the oscillation induced by the local magnetic fields $h$. To be able to trust these results, we tested the quality of this more flexible circuit on the restricted 1-dimensional slice of the landscape, which is depicted in Fig.~\ref{fig:reproduce}a (see Fig.~\ref{fig:freeangle-check} for that comparison). 

While the joint memory requirements for all 127 individually reconstructed landscapes (in our implementation) exceed the RAM capabilities of conventional laptops, re-evaluation of one snapshot in time only takes a few seconds if all landscapes are loaded into a memory. Using distributed computing resources paired with LOWESA's natural parallelizable workflow, we believe that the size and complexity of systems that can realistically be simulated could be substantially increased.

Finally, we push our algorithm to its current limit by opening up all angles for the $L=20$ Trotter steps case in Fig.~\ref{fig:4b_and_more}. This circuit contains 5420 free parameters, which for us results in $m=5420$ RZ-gates and 5760 CNOT gates in the circuit in a 2D entangling topology (which are more than twice as many RZ gates and CNOT gates as in Fig.~\ref{fig:4b_and_more}). Fig.~\ref{fig:freeangle-check} in Appendix~\ref{apx:verification} estimates the quality of the reconstructed landscape by comparing to the correlated-angle slice of the landscape with $J^{(i,j)}=J=-\pi/2$. While our results for the fully open-angle circuit are arguably of equal (or higher quality) than some of the methods presented in Refs.~\cite{kim2023evidence} and~\cite{anand2023classical}, we do not feel confident that finer details in the full landscape can be trusted. Even so, LOWESA's performance on this quantum circuit of elevated difficulty highlights its scalability relative to more established methods such as tensor networks. More computing resources, smarter truncation, or a better tree-search algorithm than the parallelized depth-first search~\cite{rao1987parallel} employed here would further benefit our method.

\section{Discussion}
In this work, we present LOWESA as an algorithm to classically construct a surrogate for the expectation landscapes in quantum simulation tasks. While this algorithm was initially introduced for simulating noisy circuits (where it enjoys efficiency guarantees), we show that it is fully capable of predicting close to exact expectation values in 127-qubit systems with thousands of entangling gates.
While conventional classical algorithms simulate quantum circuits with one fixed parameter vector $\bm\theta$ at a time, LOWESA approximately reconstructs the entire expectation landscape spanned by the circuit parameters. This process is more costly than calculating a single expectation value, but the surrogate landscape can then be re-evaluated at different parameters at a significantly lower computational cost than existing methods.

We apply this algorithm to the heavy-hex TFI system studied in Ref.~\cite{kim2023evidence} and find that it produces competitive results on a laptop. Where exact calculation is possible, LOWESA demonstrably gives high-quality predictions. Even beyond exact verification, our results are in good agreement with the state-of-the-art classical simulation methods presented in Refs.~\cite{tindall2023efficient,kechedzhi2023effective,beguvsic2023fast,torre2023dissipative,liao2023simulation, beguvsic2023converged}. Interestingly, by use of the efficiently calculable Clifford circuit boundary cases, we find that two of the five observables considered allow for trivial solutions with $\sin^{25}$ and $\sin^{34}$ for Figs.~\ref{fig:reproduce}c and \ref{fig:4b_and_more}a, respectively. These trivial solutions are however not sufficient to construct an accurate surrogate outside of the presented parameter domain. This highlights that full reconstruction of the expectation landscape, as done in our work, is vastly more challenging than the setup considered in Ref.~\cite{kim2023evidence} and follow-up works~\cite{tindall2023efficient,kechedzhi2023effective,beguvsic2023fast,torre2023dissipative,liao2023simulation, beguvsic2023converged}.

Having created the surrogate expectation landscape for the 127-qubit examples with fully opened-up angles (which in its current form took two days on a laptop), we can freely tune the initial state of the simulation, the Hamiltonian coefficients, and even the target observable. This allows us to study single-site magnetization in the system evolving in time with homogeneous or in-homogeneous couplings between sites. Evaluating one snapshot in time for all 127 observables and a new initial state only takes a few seconds on a laptop once the surrogate is loaded into memory. We stress that as well as providing a powerful tool to quickly simulate a range of physical problems, LOWESA (in contrast to many other simulation methods) is not directly limited by circuit topology.

Going forward, we believe that LOWESA can fulfill a valuable function that is complementary to existing classical simulation methods as well as quantum hardware experiments. Its ability to quickly and broadly scan expectation landscapes is a feature that no other popular simulation method shares. The surrogate landscape can, for example, be used to verify the existence and narrow down the location of critical points, to perform meta-learning of Hamiltonians with certain time evolution properties, or it can be used for tasks where explicit access to the Fourier spectrum of an observable is required (for example, phase estimation ~\cite{somma2002simulating}, or calculating Green's functions~\cite{gomes2023computing}).

How far this method can be pushed to larger and more complicated systems remains to be studied and likely requires further development of truncation and approximation techniques. We also expect symmetries in the system or quantum circuit to help find or trim paths~\cite{fontana2022non}, whose contribution can be predicted with prior knowledge. However, even in its current form presented in this work, we show that LOWESA is competitive with leading classical simulation methods in producing high-quality expectation estimates, with a likely computational scaling advantage as the number of expectation function evaluations increases. 

\section{Acknowledgements}
The authors would like to acknowledge Cristina C\^{i}rstoiu for helpful discussions. The authors would like to acknowledge Zach Morell and Carleton Coffrin for their contribution and valuable suggestions on the current implementation of the LOWESA algorithm.
This work was supported by the U.S. Department of Energy (DOE) through a quantum computing program sponsored by the Los Alamos National Laboratory (LANL) Information Science \& Technology Institute. M.S.R. was partially supported by the U.S. DOE, Office of Science, Office of Advanced Scientific Computing Research, under the Accelerated Research in Quantum Computing (ARQC) program. L.C. acknowledges support by the Laboratory Directed Research and Development (LDRD) program of LANL under project number 20230049DR. 
E.F. acknowledges support from the UK government department for Business, Energy and Industrial Strategy through the UK National Quantum Technologies Programme, and of an industrial CASE studentship, funded by the UK Engineering and Physical Sciences Research Council (grant EP/T517665/1), in collaboration with the University of Strathclyde, the National Physical Laboratory, and Quantinuum. E.F. also acknowledges support by JPMorgan Chase \& Co. via the Quantum Computing Summer Associate Program.
Z.H. acknowledges support from the Sandoz Family Foundation-Monique de Meuron program for Academic Promotion.

\bibliography{quantum,mybib}

\clearpage

\appendix

\section{Convergence with truncation parameters}\label{apx:convergence}
\begin{figure}[t]
    \centering
    \includegraphics[width=0.95\linewidth]{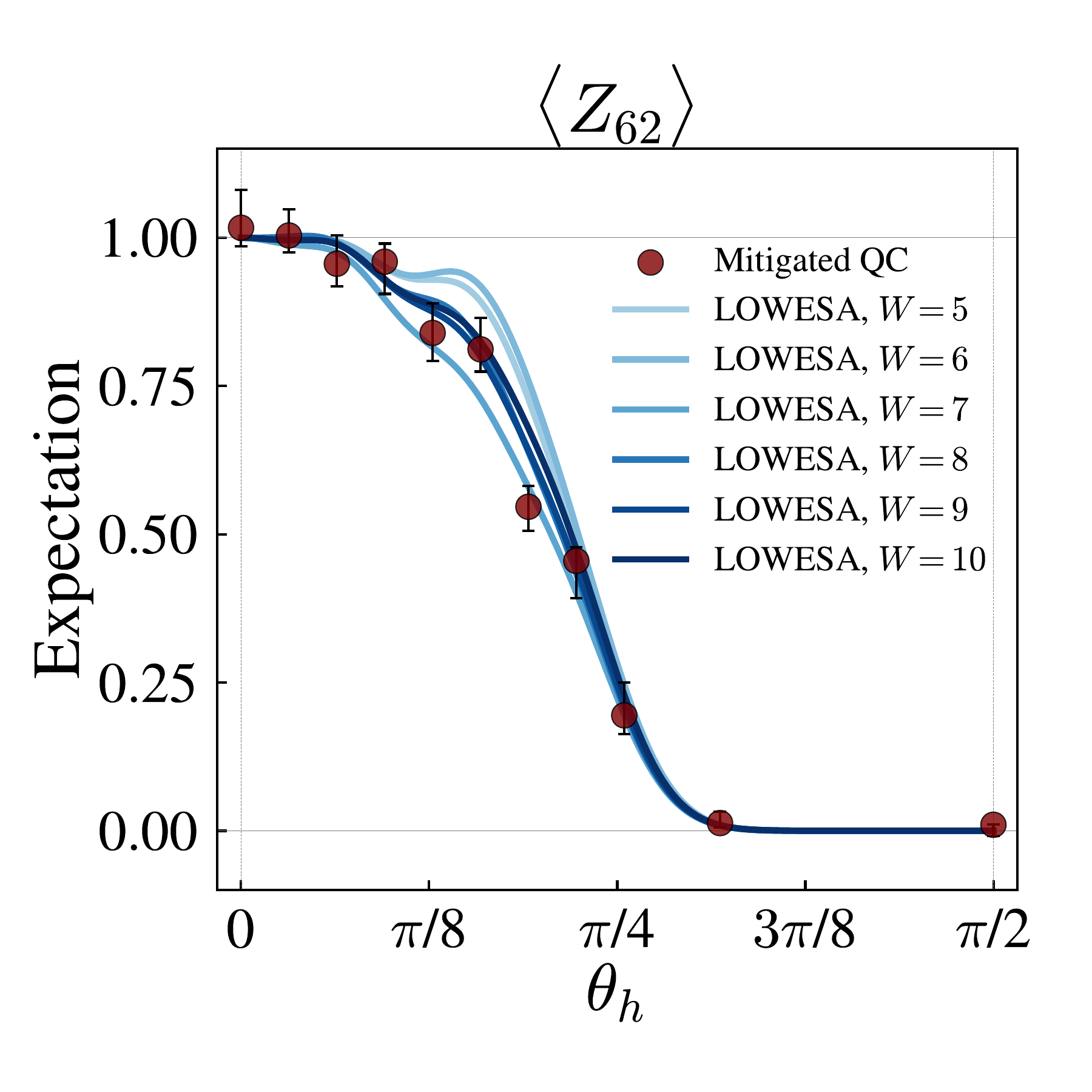}
    \caption{\textbf{Improving approximation with higher truncation weight $W$.} For the system considered in Fig.~\ref{fig:4b_and_more}a, we show the convergence of the expectation curve, and thus the surrogate landscape, with operator weight truncation $W$, given fixed values for the cutoff frequency $\ell=36$ and a truncation probability of $p=0.05$.}
    \label{fig:weight-convergence}
\end{figure}

The truncation and tailoring techniques outlined in Sec.~\ref{sec:approx_and_tailoring} can be understood as hyperparameters that control both the computational resource requirements and the quality of the surrogate landscape. Similarly to the bond dimension hyperparameter in tensor network simulations, reduction in computational cost may not come with a reduction in quality, and a more expensive simulation may not yield better results. That is, there exists a physical maximal bond dimension that allows one to capture all entanglement in the system, below which quality degrades, but above which only computational cost increases at the same quality. In that case, the simulation is said to be \textit{converged}. The same can be observed for the truncations used in this work, where, at a certain point, relaxing the truncations does not further improve performance, and one may have achieved close-to-exact results at a tractable computational cost.

There are however two significant differences between the truncations employed in this work to the well-understood bond dimension in tensor networks. 

First, the computational resource scaling of LOWESA is generally exponential in some of the current truncation parameters. Increasing the maximum frequency $\ell$ by one up to doubles the number of paths explored, and similarly increasing the maximum operator weight $W$ by one roughly doubles the chance of an operator $\textbf{U}^\dagger_{\bm\omega}|O\rrangle$ to annihilate against the initial state, i.e., $\llangle 0|\textbf{U}^\dagger_{\bm\omega}|O\rrangle = 0$. This is in contrast to the bond dimension, which generally needs to scale exponentially to faithfully simulate a quantum system, but the computational scaling is polynomial. This tends to cause discrete jumps in LOWESA's quality for every increase in $\ell$ and $W$, but convergence is still possible.

As an example, in Fig.~\ref{fig:weight-convergence} we show how the curve corresponding to the $L=20$ Trotter step system in Fig.~\ref{fig:4b_and_more} converges with increasing $W$. We recall that the operator weight truncation can be viewed as enforcing an effective entanglement lightcone per path. Thus the convergence can be attributed to the truncated lighcone approaching the true physical lightcone. On a side note, LOWESA naturally respects the entanglement lightcone induced by the circuit topology, because RZ-gates outside the entanglement lightcone never cause the paths to split.

In the version of LOWESA presented in this work, we utilized two main truncation parameters, $\ell$ and $W$. 
This multitude of truncation parameters is a second key difference between LOWESA and tensor networks, where only one truncation (i.e, bond dimension) is generally used. When LOWESA is used with multiple truncation parameters, each parameter can only converge with respect to the limits imposed by the other truncations. While the curves in Fig.~\ref{fig:weight-convergence} appear to be converging with $W$, they are still not exact because the chosen value of $l$ is too low. In such cases it may occasionally turn out that more restrictive values of a given truncation parameter (given constraints from other fixed truncation parameters) are actually closer to the true dynamics. This is seen in Fig.~\ref{fig:weight-convergence} where $W=9$ appears slightly closer to the true dynamics than $W=10$.

One may wonder why the truncation probability parameter $p$ is not mentioned above as one of the main truncation parameters. That is because in effect it acts as a relaxation of $\ell$. Assuming that the splitting probability is constant along any given path, the probability of eventual truncation can be estimated as $P(|\bm\omega|_1 \leq \ell \, | \, l, k) =  \sum_{i=0}^{m-k} \binom{m-k}{i} (\frac{l}{k})^i (1-\frac{l}{k})^{m-k-i}$, where $l$ is the number of splits so far when encountering the $(k+1)^{\text{th}}$ RZ-gate out of $m$ in total. $P$ tends to be highest for the paths that eventually end up with close to $\ell$ splits. Thus $p$ effectively reduces the contribution of frequencies close to the cutoff which are most likely to have $P(|\bm\omega|_1 \leq \ell \, | \, l, k) < p$.

It is vital to the reliability of LOWESA that future work investigates and quantifies the precise impact of individual truncations on the surrogate landscape, and potentially reveals novel truncations which have a more predictable computational cost to quality trade-off. While that is plausibly impossible for the simulation of noise-free systems in the worst case, practical cases may allow powerful heuristics.

\section{Verification of the free-angle surrogate}\label{apx:verification}
\begin{figure*}[t]
    \centering
    \includegraphics[width=0.45\linewidth]
    {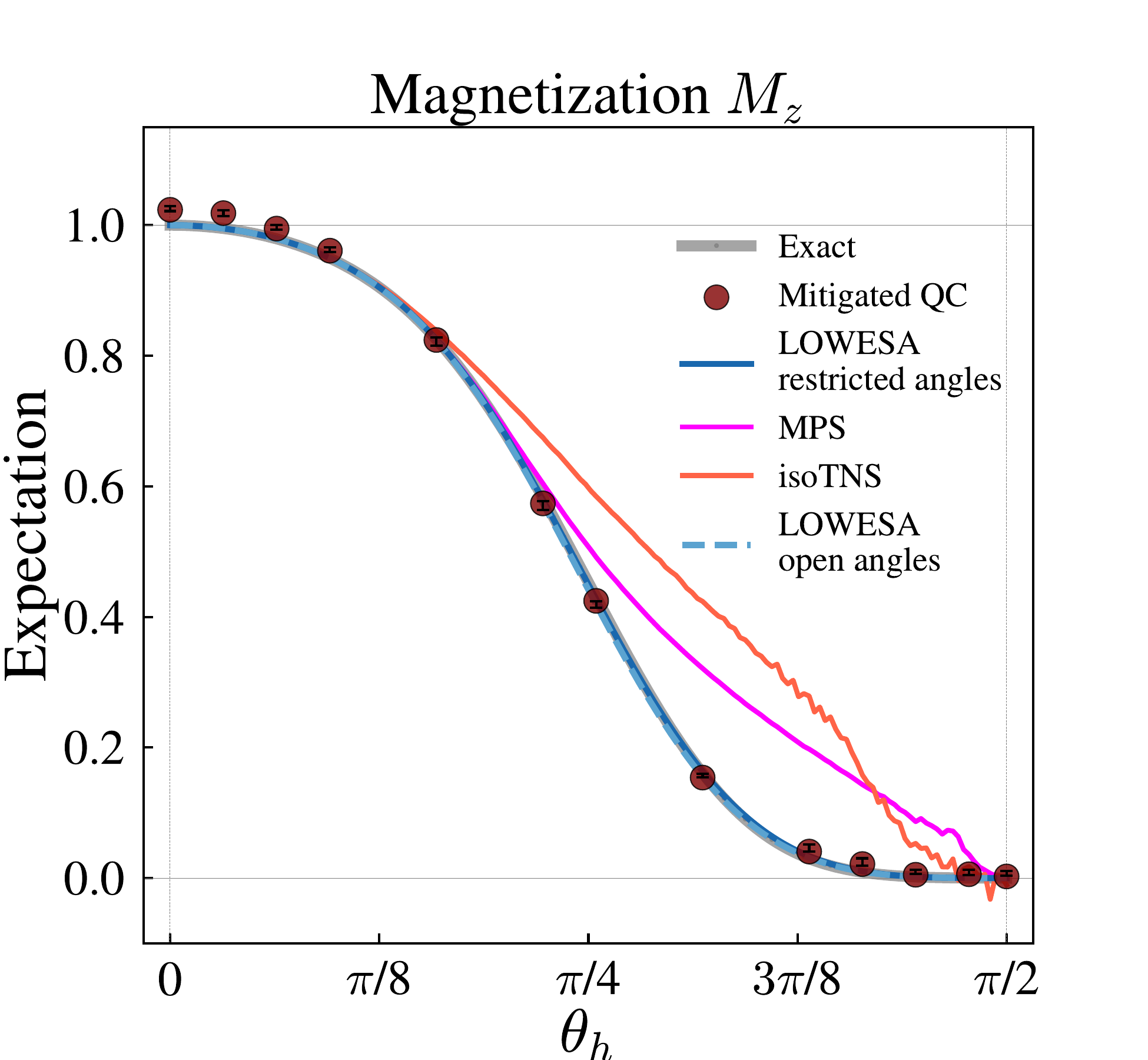}
    \includegraphics[width=0.45\linewidth]{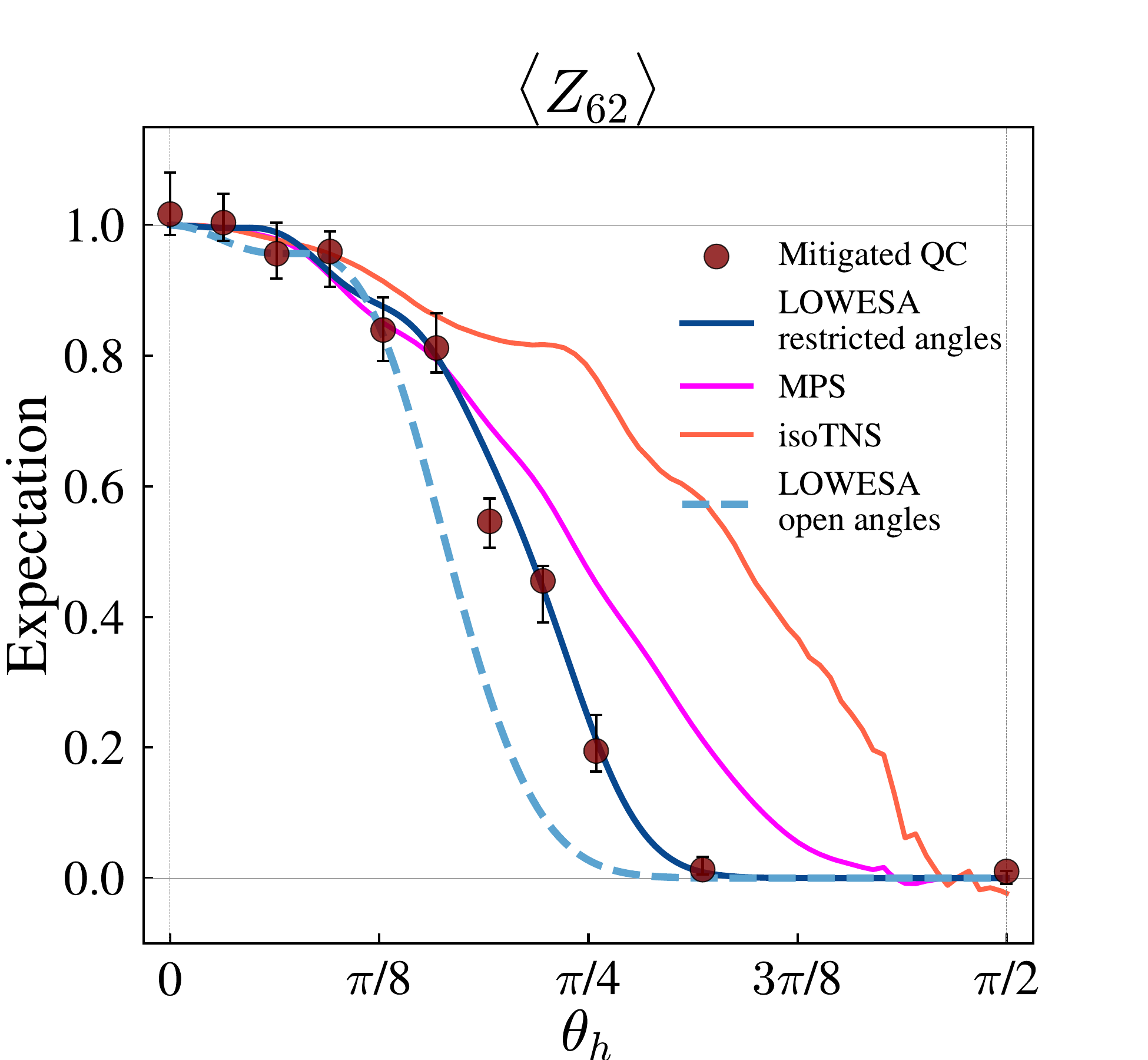}
    \caption{\textbf{Verifying the open-angle quantum circuits.} We depict the expectation curves corresponding to Figs.~\ref{fig:reproduce}a  (left) and~\ref{fig:4b_and_more}a (right), as well as the expectations produced by the surrogate using the significantly more challenging circuit in Eq.~\eqref{eq:trotter-circuit} with fully individually controllable angles. The truncation for the open-angle cases are $\ell=30, p=0.05, W=8$ (left) and $\ell=40, p=0.05, W=8$ (right).  The MPS and isoTNS simulations of the simpler restricted-angle presented in Ref.~\cite{kim2023evidence} are depicted in pink and red. 
    }
    \label{fig:freeangle-check}
\end{figure*}

In this section, we validate the performance of the fully open-angle quantum circuits discussed towards the end of Sec.~\ref{sec:Results} and presented in Fig.~\ref{fig:magnetization}. As a reminder, while the quantum circuit for the Trotter time evolution of the TFI Hamiltonian is given in Eq.~\eqref{eq:trotter-circuit}, the example benchmarks introduced in Ref.~\cite{kim2023evidence} fixed all coupling strengths $J^{(i,j)} = J = -\frac{\pi}{2\Delta t}$, such that $\theta_J^{(i,j)}=\theta_J = -\frac{\pi}{2}$. This makes the $R_{ZZ}$ entangling gates Clifford-gates (and so more efficient to classically simulate) and only require one CNOT gate to implement on IBM's quantum device~\cite{kim2023evidence}.

The quantum circuit used in Fig.~\ref{fig:magnetization} on the other hand employs a more challenging version of the circuit where all gates are freely parametrized, i.e. the circuit is actually that in Eq.~\eqref{eq:trotter-circuit}. Seeing as the surrogate landscape constructed by this circuit includes the circuit for Figs.~\ref{fig:reproduce} and~\ref{fig:4b_and_more} as a special case, we verify the quality of the landscape on those restricted slices of the landscape. 

The results for the magnetization  $M_z = \frac{1}{127}\sum_i\langle Z_i\rangle$ (see Fig~\ref{fig:reproduce}a) and the most challenging case of the $\langle Z_{62}\rangle$ observable with  $L=20$ Trotter-layers (see Fig.~\ref{fig:4b_and_more}) are shown in Fig.~\ref{fig:freeangle-check}.
We see that the expectation curve for the magnetization is nearly exact, which suggests that the surrogate landscape is overall of high quality. On the other hand, the surrogate for the $Z_{62}$ observable cannot be expected to be highly accurate for all angles since it does not match the experimental and simulation data for the simplified circuit. However, the general trend of the curve can still be argued to be of equal or better quality than some of the methods presented in Refs.~\cite{kim2023evidence} and~\cite{anand2023classical}, which highlights the scalability of the LOWESA algorithm, even on this significantly harder quantum simulation task.

\subsection*{Disclaimer}
This paper was prepared for informational purposes with contributions from the Global Technology Applied Research center of JPMorgan Chase \& Co. This paper is not a product of the Research Department of JPMorgan Chase \& Co. or its affiliates. Neither JPMorgan Chase \& Co. nor any of its affiliates makes any explicit or implied representation or warranty and none of them accept any liability in connection with this paper, including, without limitation, with respect to the completeness, accuracy, or reliability of the information contained herein and the potential legal, compliance, tax, or accounting effects thereof. This document is not intended as investment research or investment advice, or as a recommendation, offer, or solicitation for the purchase or sale of any security, financial instrument, financial product or service, or to be used in any way for evaluating the merits of participating in any transaction.

\end{document}